\documentstyle[a4, epsf,cite,12pt]{article}
\newcommand{\sumA}{\sum_{i=1}^A}
\newcommand{\sul}{${}^{32}$S}
\newcommand{\boldr}{\mbox{\boldmath$r$}}

\newcommand{\Ge}{$^{64}$Ge}
\newcommand{\Se}{$^{68}$Se}
\newcommand{\Kr}{$^{72}$Kr}
\newcommand{\Sr}{$^{76}$Sr}
\newcommand{\Zr}{$^{80}$Zr}
\newcommand{\Mo}{$^{84}$Mo}
\newcommand{\Yzero}{$Y_{30}$}
\newcommand{\Yone}{$Y_{31}$}
\newcommand{\Ytwo}{$Y_{32}$}
\newcommand{\Ythree}{$Y_{33}$}

\begin{document}

\title{
{\large\bf Symmetry-Unrestricted Skyrme-Hartree-Fock-Bogoliubov
 Calculations for Exotic Shapes in $N=Z$ Nuclei 
 from \Ge~ to \Mo}
}

\author{M. Yamagami$^1$, K. Matsuyanagi$^1$ and M. Matsuo$^2$\\
{\small\it $^1$ Department of Physics, Graduate School of Science,}\\
{\small\it Kyoto University, Kyoto 606-8502, Japan}\\
{\small\it $^2$ Graduate School of Science and Technology, }\\
{\small\it Niigata University, Niigata 950-2101, Japan}
}
\date{}
\maketitle

\begin{abstract}

By performing fully 3D symmetry-unrestricted Skyrme-Hartree-Fock-Bogoliubov 
calculations, we discuss shape coexistence  
and possibility of exotic deformations simultaneously breaking 
the reflection and axial symmetries in proton-rich $N=Z$ nuclei: 
\Ge, \Se, \Kr, \Sr, \Zr\ and \Mo. 
Results of calculation indicate that the oblate ground state 
of \Se\ is extremely soft 
against the \Ythree\ triangular deformation, 
and that the low-lying spherical minimum 
coexisting with the prolate ground state in \Zr\ is extremely soft 
against the \Ytwo\ tetrahedral deformation. 
\end{abstract}



\section{Introduction}

The Hartree-Fock-Bogoliubov (HFB) method with the Skyrme interactions
is one of the standard approaches in nuclear structure research\cite{RS80,AF90}. 
In the last two decades it has become possible to solve the HFB
equations directly in the coordinate mesh space\cite{DF84,Bu80}. 
In recent years, in order to investigate the structure of drip-line
nuclei, the need for such coordinate-space HFB calculations 
has been greatly increased and intensive analyses have been made
for neutron radii and skins in spherical neutron-rich nuclei
\cite{SD93,DN96a,DN96b,MD00,RB97,Ta98,Ta00}: 
Since the easier HF plus BCS method
breaks down when treating the pairing correlation in weakly bound
systems due to a leakage of nucleons into the continuum,
we need to calculate the mean-field (particle-hole) correlations
and the pairing (particle-particle) correlations selfconsistently
in order to preserve confinement of the nuclear density while allowing
the pairing excitations to positive energy resonant states\cite{DF84}
(see, e.g. \cite{Me98} and references therein for mean-field approaches
other than the Skyrme-HFB method).        

Recently, Terasaki, Heenen, Flocard and Bonche\cite{TH96,TF97} 
have removed the restriction of spherical symmetry in solving the
coordinate-space Skyrme-HFB equations in order to investigate
the possibility to get three-dimensional (3D) deformed solutions
in neutron rich nuclei. In their works,
a Skyrme interaction is used to describe the Hartree-Fock (HF)
Hamiltonian while a density dependent zero-range interaction is used
for the pairing channel. The mean-field HF equations are solved by the
imaginary-time evolution method\cite{DF80} 
in a 3D cubic mesh space while the HFB equations are
solved in terms of the two-basis method 
developed earlier in \cite{GB94,TH95}.
The discretization in 3D mesh space has the advantage 
over methods relying on an expansion in the harmonic-oscillator basis 
that nuclei with exotic deformations can be treated at the same level 
of accuracy\cite{FH82,BF85,BF87}.  
In these works, however, reflection symmetries with respect to three
planes are imposed for the nuclear density so that only one spatial
octant is needed to solve the HFB equations.

The major purpose of this paper is to extend their method by removing
the symmetry restrictions mentioned above and investigate the
possibility of exotic shapes simultaneously breaking the axial and reflection
symmetries in the mean field. 
For this purpose, we have constructed a new computer code that carries out 
Skyrme-HFB calculations in the 3D Cartesian-mesh space 
without imposing any restrictions on the spatial symmetry.
Recently, on the basis of the Skyrme HF plus BCS calculations
with no restriction on the nuclear shape, 
Takami, Yabana and Matsuo\cite{TY98,MT99}
suggested that the oblate ground state of \Se~
is extremely soft against the \Ythree~ triangular deformation, and 
that the low-lying ``spherical'' minimum coexisting with the prolate ground 
state in \Zr~ has the \Ytwo~ tetrahedral shape. 
As the first application of a fully 3D, symmetry-unrestricted Skyrme HFB 
method with the use of the density-dependent, zero-range pairing 
interaction\cite{TH96,TF97,TH95,Ch76,BE91,SZ92,TB93,ZR87,RZ88,FT94,BF96}, 
we investigate in this paper shape coexistence 
and possibility of non-axial octupole deformations in
proton rich $N=Z$ nuclei in the $A=64-84$ region and examine the
above predictions. These nuclei are especially interesting objects to
study, since proton and neutron deformed shell effects act coherently and 
rich possibilities arise for coexistence and competition of 
different shapes (see \cite{EM88} for earlier references). 
In recent years, active experimental studies of these nuclei are going on
by means of combinations of radioactive nuclear beams and new gamma-ray and
charged particle detector systems (see \cite{Si99,Na99,Mu99,Ta99} for reviews).
It should be noted here that, although extensive theoretical calculations and
rich experimental evidences have been accumulated for axially symmetric
octupole (\Yzero) deformations, as reviewed in \cite{AB93,BN96},
only a few calculations using Woods-Saxon Strutinsky methods are
available \cite{Sk91,Ch91,LD91,LD94} except for light nuclei, 
and no firm experimental evidence exists up to now 
concerning the non-axial octupole (\Yone, \Ytwo, \Ythree) deformations 
in the mean fields. 
For light nuclei, non-axial octupole deformations have been
discussed\cite{TY96,LL75,EF70,OS71,EE85} in connection 
with alpha-cluster structures\cite{FH80};
for instance, a triangular structure of $^{12}$C\cite{TY96,EF70}
and a tetrahedral shape for $^{16}$O\cite{OS71,EE85}.

Our motive for developing the  coordinate-space 
Skyrme-HFB method is not only to investigate the possibilities 
of emergence of new types of symmetry breakdown
in the ground states of proton-rich and neutron-rich nuclei,
but also to investigate, in the future, low-lying modes of excitation 
of such unstable nuclei by means of the RPA and 
the Selfconsistent Collective Coordinate (SCC) method\cite{MM80} 
on the basis of the HFB basis thus obtained. 
We intend to proceed in parallel with other calculations with the 
use of more phenomenological shell model potentials and separable
interactions. The Skryme-HFB method is suited for this aim, 
as it provides a local mean-field potential so that such a
comparative study is easy.

In Section 2,  a brief account of the method of the coordinate-space
Skyrme-HFB calculation is given. 
In Section 3, results of numerical calculation are presented and discussed.
In Section 4, a conclusion is given.

\section{Skyrme-HFB calculation}

\subsection{\it Two basis method}

For convenience, we here recapitulate the two basis method 
\cite{TH96,TF97,GB94,TH95} adopted as the algorithm of our computer code. 
In this method, the imaginary-time evolution method is combined with a
diagonalization of the HFB Hamiltonian matrix to construct the
canonical basis. 

We first determine the single-particle wave functions $\phi_i$\ 
satisfying the HF equations
\begin{eqnarray}
h\left[ {\rho\left( \boldr \right)} \right]
\phi_i \left( \boldr \right) = \varepsilon_i \phi_i 
\left( \boldr \right) 
\end{eqnarray}
\noindent
by means of the imaginary-time evolution method\cite{DF80}. 
Here $h$, $\varepsilon_i$ and $\rho\left( \boldr \right)$   
denote the mean-field Hamiltonian, the single-particle energies 
and the total nuclear density, respectively. 
(The isospin index $\tau$ is omitted for simplicity.)
We next diagonalize the HFB Hamiltonian matrix \cite{RS80} 

\begin{eqnarray}
\left( {\matrix{{h-\lambda }&\Delta \cr
{-\Delta ^*}&{-h^*+\lambda }\cr
}} \right)\left( {\matrix{{U_k}\cr
{V_k}\cr
}} \right)=E_k\left( {\matrix{{U_k}\cr
{V_k}\cr
}} \right) 
\end{eqnarray}
\noindent
to get the one-body density matrix $\rho$ and the pairing tensor $\kappa$:

\begin{eqnarray}
\rho =V^*V^T,\quad \kappa =V^*U^T. 
\end{eqnarray}

\noindent
We then diagonalize the density matrix $\rho$ 
and obtain the occupation coefficients $n_{\alpha}$  
and the unitary transformation $W$ which relates 
the HF wave functions $\phi_i$ to the canonical basis wave functions 
$\varphi_{\alpha}$:

\begin{eqnarray}
\rho _{kl}=\sum\limits_\alpha  {n_\alpha W_{k\alpha }
W_{\alpha l}^{\dag}} 
\end{eqnarray}

\begin{eqnarray}
\varphi _\alpha \left( \boldr \right)=\sum\limits_j 
{W_{j\alpha }\phi _j\left( \boldr \right)}. 
\end{eqnarray}
\noindent
In the canonical basis $\varphi_{\alpha}$, the HFB density matrix 
in the coordinate space is diagonal:

\begin{eqnarray}
\rho\left( \boldr, \boldr' \right)=\sum\limits_\alpha  
{n_\alpha \varphi _\alpha \left( \boldr \right)\varphi _\alpha 
\left( \boldr' \right)^*}. 
\end{eqnarray}

\noindent
These steps are repeated until the convergence is achieved. 

The single-particle wave functions and densities are represented 
on a full 3D Cartesian mesh space within a spherical container.
In the present calculation, the radius of the spherical container 
and mesh spacing are set to $R_{mesh}=10.0$ fm and $h=1.0$ fm, 
respectively. Tajima et al.\cite{TT96,TT98} 
have carefully examined possible errors due to the use of 
the mesh size $h = 1.0$ fm and they found that,
since discretization errors are essentially independent of the
nuclear shape, deformation energies obtained with this mesh size 
are quite accurate (see also \cite{HN77}). 
Actually, we have constructed the new Skyrme-HFB code by extending
the cranked Skyrme-HF code\cite{YM00} written previously and applied to the
investigation of the yrast structure of \sul, so that the cranking
term can be included. In this paper, however, we examine only the
cases of zero angular momentum.

\subsection{\it The Skyrme plus density-dependent pairing interactions} 

We use the SIII parameter set \cite{BH75} of the Skyrme interaction
for the mean-field (particle-hole) channel,  
which has been successful in describing systematically the ground
state quadrupole deformations in proton and neutron rich
Kr, Sr, Zr and Mo isotopes\cite{BF85} 
and in a wide area of nuclear chart\cite{TT96}.
For the pairing (particle-particle) channel, 
we use the density-dependent zero-range interaction
\cite{TH96,TF97,TH95,Ch76,BE91,SZ92,TB93,ZR87,RZ88,FT94,BF96},
which has been successful in describing, for instance,
the odd-even staggering effects in charge radii,

\begin{eqnarray}
V_{pair} \left( \boldr_1, \boldr_2 \right) = 
\frac{V_0}{2}\;\left( {1-\hat P_\sigma } \right)
\;\left( {1-{{\rho \left( \boldr_1 \right)} \over {\rho _c}}} \right)
\;\delta \left( \boldr_1 - \boldr_2 \right) 
\end{eqnarray}

\noindent
with the notation of \cite{TH95}, 
where the strength $V_0$ and the density $\rho_c$ are 
parameters and $\hat P_{\sigma}$ denotes the spin exchange operator. 
For these parameters, we use the standard values \cite{TH95, TF97}:
$V_0 = -1000.0$ MeV$\cdot$fm${}^3$, $\rho_c = 0.16$ fm${}^{-3}$.
The pairing interaction is smoothly cut off at 5 MeV 
above the Fermi energy in the same way as in \cite{BF85}.   
For a more general form of the density dependent pairing
interaction, we refer \cite{FZ96,FT00}.

To check the dependence on the Skyrme-interaction parameter sets,
we make calculations with the SkM$^*$ \cite{BQ82} and
SLy4 \cite{CB97} sets for an example of \Se.
We refer to a recent work by Reinhard et al.\cite{RD99} for a detailed
and systematic study of shape coexistence phenomena in relation to the
properties of various versions of the Skyrme interaction.
We shall also check the dependence on the pairing strength $V_0$ adopted.

\subsection{\it Constrained HFB calculation}

In order to investigate the deformation properties away from
the HFB equilibrium points, we perform constrained HFB calculations 
with the use of the quadratic constraints for the mass-quadrupole 
(octupole) moments \cite{FQ73} to obtain the energy surfaces as 
functions of the quadrupole (octupole) deformations. 
Because no spatial symmetry is imposed on the 3D mesh space, 
the center of mass and the directions of the principal axes of 
the nucleus can move freely without affecting the total energy. 
To evaluate the physical quantities like deformation parameters, 
it is crucially important to fulfill the constraints 
to keep the center of mass,  

\begin{eqnarray}
\left\langle \sumA x_i \right\rangle =\left\langle \sumA y_i \right\rangle 
=\left\langle \sumA z_i \right\rangle =0, 
\end{eqnarray}
\noindent
and the directions of the principal axes, 

\begin{eqnarray}
\left\langle \sumA (xy)_i \right\rangle =\left\langle \sumA (yz)_i \right\rangle 
=\left\langle \sumA (zx)_i \right\rangle =0. 
\end{eqnarray}

These requirements are taken care of by means of the quadrupole constraints 
on these conditions as in our previous study \cite{YM00}.  

\subsection{\it Deformation parameters} 

As measures of the deformation, we calculate the mass-multipole moments,  

\begin{eqnarray}
\alpha _{lm}={{4\pi } \over {3AR^l}}\int r^l X_{lm}\left( 
{\Omega} \right)\rho \left( \boldr \right)d \boldr
,\quad (m=-l,\cdots ,l)
\end{eqnarray}

\noindent
where 
$\rho \left( \boldr \right)=\sum\limits_\alpha  
{v_\alpha ^2\left| {\varphi _\alpha \left( \boldr \right)} \right|^2}$,
$R=1.2 A^{1/3}$ fm and 
$X_{lm}$ are real bases of the spherical harmonics,

\begin{eqnarray}
	X_{l0} & = & Y_{l0},
		 \\
	X_{l|m|} & = & \frac{1}{\sqrt{2}}( Y_{l-|m|}+Y_{l-|m|}^{*} ),
		   \\
	X_{l-|m|} & = & \frac{-i}{\sqrt{2}} ( Y_{l|m|}-Y_{l|m|}^{*} ).
\end{eqnarray}
\noindent
Here the quantization axis is chosen as the largest (smallest) 
principal axis for prolate (oblate) solutions. 
We then define the quadrupole deformation parameter $\beta_2$,
the triaxial deformation parameter $\gamma$, and the
octupole deformation parameters $\beta_3$ and $\beta_{3m}$ by

\begin{eqnarray}
\alpha _{20}=\beta _2\cos \gamma ,
\quad \alpha _{22}=\beta _2\sin \gamma,
\end{eqnarray}
\begin{eqnarray}
\beta _3=\left( {\sum\limits_{m=-3}^3 {\alpha _{3m}^2}} \right)^{1/2},
\quad \beta _{3m}=\left( {\alpha _{3m}^2+\alpha _{3-m}^2} \right)^{1/2}
\quad \left( {m=0,1,2,3} \right). 
\end{eqnarray}

For convenience, we also use the familiar notation $-\beta_2$ for
oblate shapes with $(\beta_2,\gamma=60^\circ)$.

\section{Results and discussions}

\subsection{\it Quadrupole deformations} 

The solutions of the Skyrme-HFB equations 
obtained in the numerical calculations for
\Ge, \Se, \Kr, \Sr, \Zr\ and \Mo\ are summarized in Table 1.
The calculated ground-state shape changes from triaxial (\Ge), 
oblate (\Se, \Kr), large prolate (\Sr, \Zr), 
to spherical shape (\Mo) with increasing $N (=Z)$. 
For \Se, \Kr, \Sr, \Zr\ and \Mo, we obtain two or three local minima 
close in energy, indicating shape coexistence. 
These gross features are consistent with available experimental data
\cite{EL91,LE90,SF98,FB00,AF97,LC87,BR97}
and previous theoretical calculations
\cite{BF85,TY98,MT99,TT96,HM84,ND85,DN87,MG92,KF93,LS95,PS96,PS00,HM96,
HZ95,SM99}.

\begin{figure}
\epsfxsize=13cm
\centerline{\epsffile{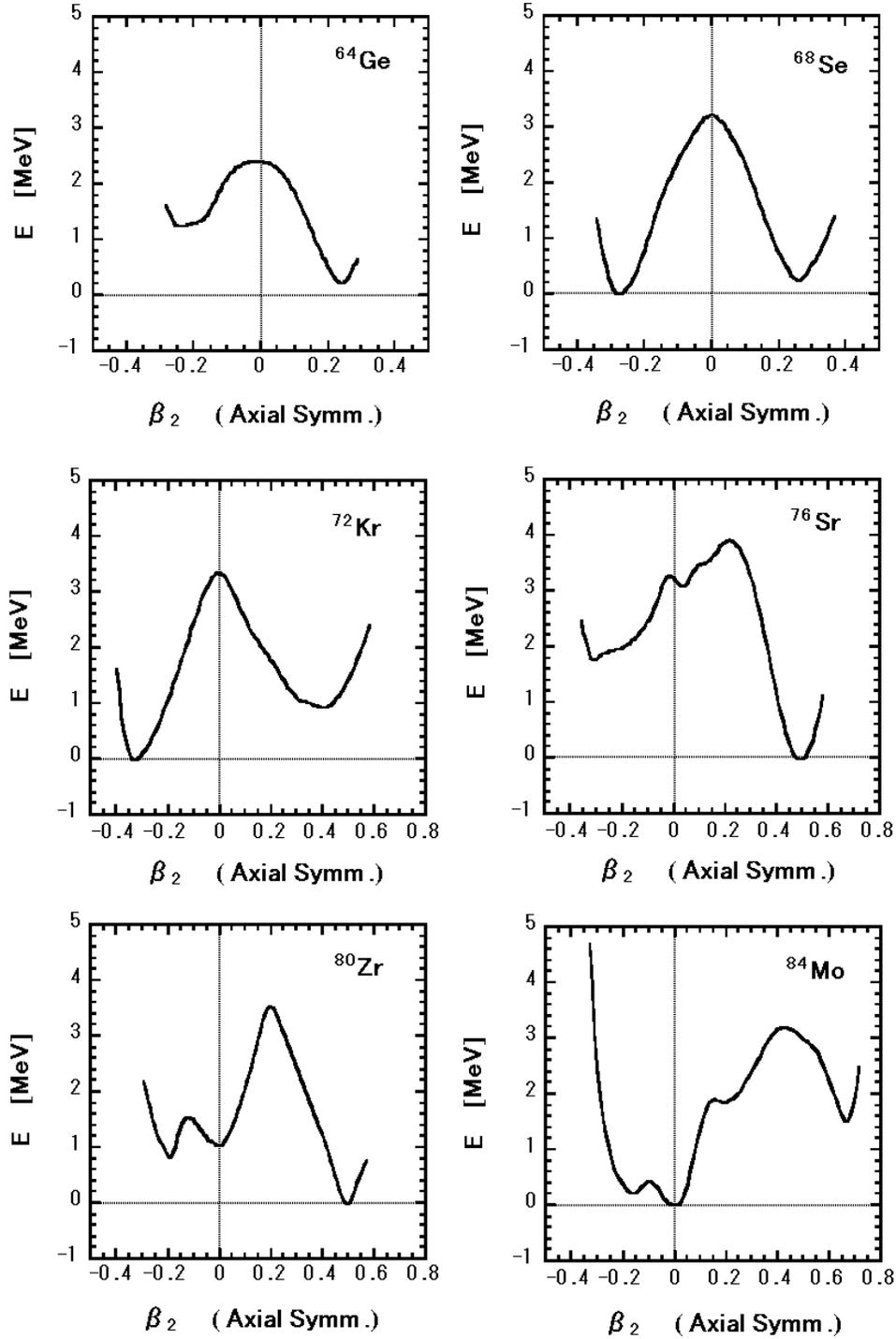}}
\caption{
\small Potential energy curves calculated by the constrained 
Skyrme-HFB method for \Ge, \Se, \Kr, \Sr, \Zr\ and \Mo\ 
are drawn as functions of the quadrupole deformation parameter $\beta_2$.
The SIII interaction is used for the particle-hole channel,
while the density-dependent pairing interaction with
$V_0 = -1000.0$ MeV$\cdot$fm${}^3$ and $\rho_c = 0.16$ fm${}^{-3}$
is used for the particle-particle channel.
}
%\vspace*{10pt}
\label{B2-E}
\end{figure}

\begin{figure}
\epsfxsize=13cm
\centerline{\epsffile{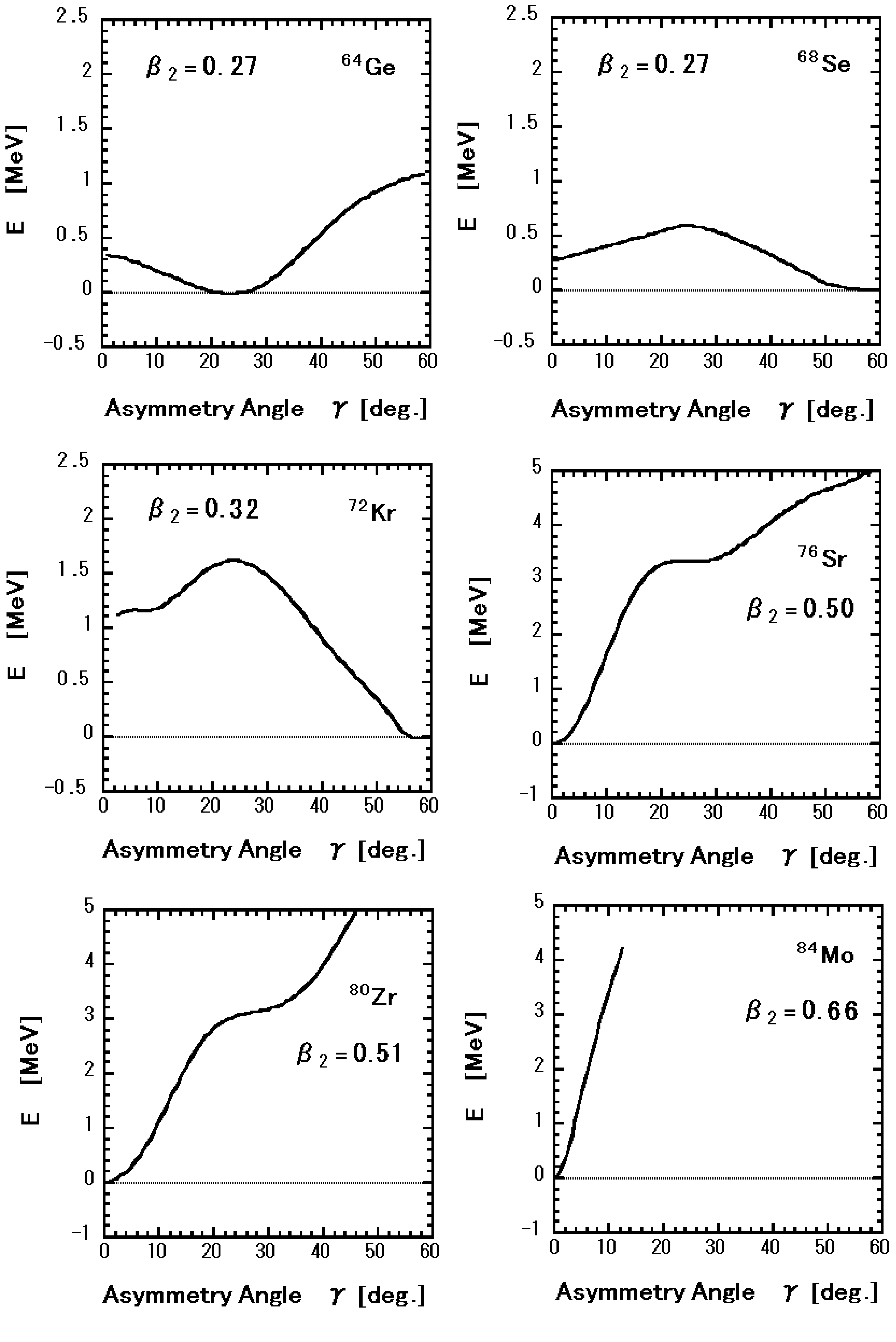}}
\caption{
\small Potential energy curves calculated at fixed $\beta_2$
by the constrained Skyrme-HFB method for \Ge, \Se, \Kr, \Sr, \Zr\ and \Mo\  
are drawn as functions of the triaxial deformation parameter $\gamma$. 
The effective interactions used are the same as in Fig. \ref{B2-E}.
}
%\vspace*{10pt}
\label{GAM-E}
\end{figure}

The potential energy curves obtained by the constrained HFB calculations
are displayed in Fig. \ref{B2-E} as functions of the quadrupole deformation
parameter $\beta_2$ and in Fig. \ref{GAM-E} as functions of the
triaxial deformation parameter $\gamma$.
Below we remark on some specific points.

As seen in Fig. \ref{GAM-E}, the calculated potential energy curve
for \Ge\ is rather shallow with respect to the $\gamma$ degree of freedom 
so that this nucleus may be regarded as "$\gamma$-soft."
This result is consistent with the experimental indication\cite{EL91}
and also with the shell model calculation by the Monte Carlo diagonalization 
method\cite{HM96}. 

Quite recently, an excited prolate band coexisting with 
the ground-state oblate band has been found 
in \Se\ \cite{FB00}. Their quadrupole deformations are estimated as
$\beta_2 \approx 0.27$ and $\beta_2 \approx -0.27$, respectively.
Although the prolate excited band-head $0^+$ state has not yet been observed, 
its excitation energy is estimated to be about 0.6 MeV.
Our calculated energy difference between the prolate and the oblate HFB
solutions, 0.52 MeV, is in good agreement with this experimental data.
The barrier between the prolate and the oblate minima is about 3 MeV in the
plot with respect to $\beta_2$ in Fig. \ref{B2-E}, but it is only about
0.3 MeV in the plot with respect to the triaxial deformation parameter $\gamma$
in Fig.\ref{GAM-E}. It might be considered that, if the barrier is so low,
the two bands built on the prolate and the oblate solutions interact strongly
so that the shape coexistence picture is too much perturbed in contradiction with
the experiment\cite{FB00}.
In our view, however, description of dynamics by going beyond the static mean-field 
approximation is necessary in order to discuss the interaction between 
the oblate and the prolate structures. In any case, understanding this
shape coexistence dynamics is an interesting subject for future.

The second minimum with $\beta_2 \approx 0.66$ seen in the potential energy 
curve for \Mo\ in Fig. \ref{B2-E} may be regarded as a superdeformed solution,
since it is related to the $Z=N=42$ deformed shell gap \cite{ND85}
formed by occupying the down-sloping [431]1/2 levels from the upper major shell 
by two protons and two neutrons.
This second minimum was also obtained in \cite{TY98}.
It offers an interesting possibility that a superdeformed rotational
band might be observed at such a low excitation energy as about 1.5 MeV.
From a viewpoint of deformed shell structure, 
the ground-state solutions for \Sr\ and \Zr\ 
have characteristics different from the
second minimum in \Mo\, and may be distinguished from the superdeformation,
although they have large prolate deformations of $\beta_2 \approx 0.5$.

\subsection{\it Non-axial octupole deformations} 

\begin{figure}
\epsfxsize=13cm
\centerline{\epsffile{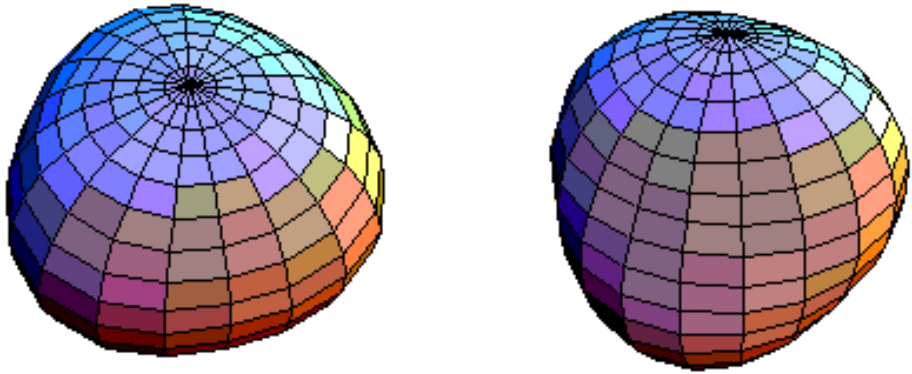}}
\caption{
\small Density contour surfaces at the half central density
of the Skyrme-HFB solution with the oblate plus triangular shape 
($\beta_2 = -0.28, \beta_{33} = 0.08$) for \Se~(left-hand side) 
and that with the tetrahedral shape 
($\beta_2 = 0.00, \beta_{32} = 0.15$) for \Zr~(right-hand side),
listed in Table 1.  
}
%\vspace*{10pt}
\label{PLOT-DEN}
\end{figure}

As a result of the Skyrme-HFB calculations for proton-rich $N=Z$ nuclei
from \Ge\ to \Mo\ (summarized in Table 1), we have found equilibrium shapes 
with finite non-axial octupole deformations for \Se\ and \Zr.
The density distribution at the HFB local minimum for \Se\ with the 
triangular deformation superposed on the oblate shape 
and that for \Zr\ with the tetrahedral deformation are illustrated in 
Fig. \ref{PLOT-DEN}.

In addition to the two cases mentioned above,
Takami et al.\cite{TY98} and Matsuo et al.\cite{MT99}
obtained, in their Skyrme-HF plus BCS calculations,
finite equilibrium values of octupole deformations  
superposed on an oblate shape in \Sr\ and also 
on a near spherical shape in \Mo. 
According to their calculations, the potential-energy curves are
very soft with respect to the octupole deformation degrees of freedom
especially in the four cases mentioned above.
In order to see the properties of the potential-energy curve in the
neighborhood of the HFB equilibrium points 
and to make a better comparison with 
the results of Refs. \cite{TY98,MT99}, we have carried out constrained HFB 
calculations with respect to the $\beta_{3m} (m=0,1,2,3)$ degrees of freedom
about the local minima (seen in Fig. 1) 
of the quadrupole deformation energies.

Figures 4, 5 and 6 show the potential-energy curves with respect to the
octupole deformation parameters $\beta_{3m}$ about the oblate, 
the spherical and the prolate (or triaxial) minima of the quadrupole 
deformation energy curves, respectively.
These curves are obtained by the constrained HFB calculations 
with the octupole operators $r^3X_{3\vert m\vert}$ as constraints.
We see that the oblate shape of \Se\ is extremely soft against the triangular
($\beta_{33}$) deformation and that the spherical shape of \Zr\ 
is extremely soft against the tetrahedral ($\beta_{32}$) deformation,
in agreement with those of the Skyrme-HF plus BCS calculations 
of Refs. \cite{TY98,MT99}.
The oblate shape of \Sr\ is fairly soft with respect to
the $\beta_{32}$ and  $\beta_{33}$ deformations and the spherical ground state
of \Mo\ is barely stable against all $\beta_{3m}$ degrees of freedom,
especially against $\beta_{30}$.
In \cite{MT99} an oblate solution with a finite equilibrium value of
$\beta_{32}$ is obtained for \Sr, while a similar solution for \Sr\
but with a finite equilibrium value of $\beta_{33}$ and
also a nearly spherical solution for \Mo\ with a finite equilibrium value of
$\beta_{30}$ is reported in \cite{TY98}.
Although such details differ depending on the treatment of the pairing 
correlations, the basic features, i.e., the softness to both $\beta_{32}$ 
and $\beta_{33}$ of the oblate shape of \Sr\ and the softess to 
$\beta_{30}$ of the spherical shape of \Mo\ are in common between the
present HFB calculations and those of \cite{TY98,MT99}.
Generally speaking, Figs. 4-6 indicate that the oblate shapes
are softer for octupole deformations $\beta_{3m}$ with higher values 
of $m$, while the prolate shapes favor lower values of $m$.

Below we focus our attention on the triangular deformation in \Se\
and the tetrahedral shape in \Zr\
and discuss about the microscopic origins of them.\\ 

\noindent
{\it Triangular deformation in} \Se\\

Generally speaking, octupole correlations are associated with strong couplings
between the shell-model orbits with $\Delta l=\Delta j=3$\cite{AB93,BN96}. 
In the $A=64-84$ region under consideration, 
they are 1g$_{9/2}$ and 2p$_{3/2}$.
In order to understand why the oblate shape in \Se\ is unstable (or extremely
soft) against the triangular deformation, 
however, we need to examine the interplay
of the quadrupole and octupole deformation effects. 
Namely, as explained below, the emergence of
the triangular deformation is strongly correlated with 
the magnitude of the oblate deformation.  

When \Se\ ($N=Z=34$) is oblately deformed,
the high $\Omega$ levels [404]$\frac{9}{2}$ and [413]$\frac{7}{2}$ 
stemming from the 1g$_{9/2}$ orbit go down in energy and approach
the Fermi surfaces for $N=Z=34$ and strong \Ythree\ couplings 
with [301]$\frac{3}{2}$ and [310]$\frac{1}{2}$ levels 
(associated with the 2p$_{3/2}$ orbit) take place. 
These \Ythree\ coupling effects are seen as repulsions between these levels
in Fig. \ref{SPE-SE} which displays the neutron single-particle energies 
as functions of the triangular deformation parameter $\beta_{33}$.
Here, the single-particle energies mean eigenvalues of the HF
Hamiltonian with  the density $\rho\left( \boldr \right)$
determined by the HFB equations, and 
the asymptotic Nilsson quantum numbers are used only for
convenience of labeling these levels:
they are, of course, not good quantum numbers.

In this figure, results of calculation 
with use of the SkM$^*$ and SLy4 interactions are also shown for comparison.
We note that the \Ythree\ coupling effects are slightly weaker 
in the case of the SkM$^*$ and SLy4 interactions in comparison with 
the case of the SIII interaction.
This is because the spacings between the levels coupled by the
\Ythree\ operator are the smallest for the SIII interaction:
The spacings at the oblate equilibrium deformations 
between the [404]9/2 and [301]3/2  levels are about 2.8, 3.4 and 3.6 MeV, 
and those between the [413]7/2 and [310]1/2  levels
are about 3.8, 4.1 and 4.2 MeV for the SIII, SkM$^*$ and SLy4 interactions, 
respectively.
Thus, as shown in Fig. \ref{SK-DEP}, the potential energy curve
with respect to the triangular $\beta_{33}$ deformation is softest 
for the case of the SIII interaction,
although they are soft also for the cases of the SkM$^*$ and SLy4 interactions.
Note that, in making this comparison, 
we have chosen the pairing-interaction strength $V_0$
such that the resulting pairing gaps $\Delta$ take about the same values for 
calculations with different Skyrme interactions (in order to make the effects
of the pairing correlations approximately the same for all cases), as shown in
the right-hand part of Fig. \ref{SK-DEP}.

The importance of the triangular \Ythree\ deformation superposed on the
oblate shape was previously pointed out by Frisk, Hamamoto and May\cite{FH94}
in terms of a two-level model as well as the modified oscillator model
which simulates the one-particle spectra in an infinite-well potential.
Our result of the Skyrme-HFB calculation provides a realistic example
which is consistent with their arguments.\\

\noindent
{\it Tetrahedral deformation in} \Zr\\

As shown by Hamamoto, Mottelson, Xie and Zhang\cite{HM91},
the tetrahedral symmetry associated with the $Y_{32}$ deformation brings about
a bunching of the single-particle levels 
and create a remarkable shell structure:
The $N=Z=40$ is one of the magic numbers for such tetrahedral shapes. 
Such a shell effect is common to various finite Fermion systems, and in fact
the tetrahedral deformation has been predicted, 
for instance, for sodium clusters
consisting of 40 atoms by the density functional Kohn-Sham 
calculation\cite{RK97,
KK98}, in which there is no spin-orbit coupling.
The instability of the spherical shape of \Zr\ against the $Y_{32}$
deformation, as exhibited in Fig. \ref{B3-E_sph}, is evidently connected 
to the magic number $N=Z=40$ for the tetrahedral shape. 

Figure \ref{SPE-ZR} shows the single-particle energy diagrams as function of 
octupole deformation parameter $\beta_{3m} (m=0,1,2,3)$. 
As expected, we can see for the case of $m=2$ a remarkable bunching of 
single-particle levels and an increase of the shell gap
at $N=40$ with increasing $\beta_{32}$,
while the other octupole deformations $(m=0,1,3)$ do not exhibit
such a feature.
Looking into details, one notices a fine splitting of the 1g$_{9/2}$ level
into three levels which correspond to
irreducible representations of the double tetrahedral 
(spinor-$T_d$) group\cite{OS71,LD94}; a twofold-degenerate level and
two fourfold-degenerate levels.

Thus, the tetrahedral shell gap at $N=Z=40$ emerges even under the
presence of the strong spin-orbit coupling. It should also be noted that the
tetrahedral minimum is obtained in the calculation selfconsistently
including the pairing correlations.

\subsection{\it Pairing gaps} 

In this subsection, we first examine dependence 
of the pairing gaps on deformations, and then discuss dependence of 
the non-axial octupole deformations on the pairing strength.
The result of calculation for the pairing gaps at equilibrium deformations 
in each nucleus is listed in Table 1.
As the pairing gaps in the HFB theory depend on single-particle levels,
the numbers listed in this Table are averages of the diagonal
elements in the HF basis, $\Delta_{i\bar{i}}$, over 5 MeV interval 
in the vicinity of the Fermi surfaces. 

In the literatures, slightly different quantities like
averages of the diagonal matrix elements in the canonical basis, 
$\Delta_{\alpha\bar{\alpha}}$, weighted by the coefficients 
of the Bogoliubov transformation, $u_\alpha v_\alpha$\cite{TO94,BR00,DB00}
or $v_\alpha^2$\cite{DF84}, are used for similar purposes.
Figure \ref{DELTA-DEF} compares these quantities for the case of
triangular deformations superposed on the oblate shape in \Se.
We see that the two average quantities, $\langle\Delta_{i\bar{i}}\rangle$ and
$\langle\Delta_{\alpha\bar{\alpha}}u_\alpha v_\alpha\rangle$, 
are approximatelty equal. 
We also confirm that the averages do not significantly depend on the
averaging interval.

Figures \ref{B2-DELTA}, \ref{GAM-DELTA} and \ref{B3-DELTA} 
display the variation of the pairing gaps 
with the quadrupole deformation parameter $\beta_2$,
the triaxial deformation parameter $\gamma$, and the
octupole deformation parameters $\beta_{3m} (m=0,1,2,3)$, respectively. 
We observe that gross features of deformation dependence of the
pairing gap correlate with 
the corresponding potential-energy curves displayed in
Figs. 1, 2 and 4-6.
Such correlations are rather easy to be understood from the
behavior of the single-particle level density near the Fermi surface,
i.e., from the well-known (spherical or deformed) shell effects
that the level density near the Fermi surface becomes relatively 
low in the vicinities of the local minima of the potential energy curve
\cite{BD72}.
Thus, the pairing correlation becomes weaker and the paring gap
decreases near the local minima. 
On the other hand, the level density becomes relatively high and
the pairing gap increases near the local maxima of the potential-energy curve. 

Because of significant shape changes in the sequence of isotopes (isotones)
in the $A=64-84$ region,
it is not always easy to extract the magnitudes of pairing correlations 
from experimental odd-even staggerings of binding energies and to assess
the appropriateness of the pairing-interaction strength 
$V_0=-1000$ MeV$\cdot$ fm$^3$ used in our HFB calculations.
Quite recently, however, Satu\l a, Dobaczewski and Nazarewicz\cite{SD98} 
have proposed a method for separating out the pairing correlation effects 
from the deformed mean-field (single-particle energy) effects 
on the odd-even staggerings, and evaluated average pairing gaps; 
these are in the range $1.0-1.6$ MeV 
for the mass region under consideration\cite{DM00}.
We note that these values agree rather well with the 
well-known global trend $\bar{\Delta}=12/\sqrt A$ MeV\cite{BM69}, 
which are in the range $1.3-1.5$ MeV for $A=64-84$.
Our calculated values of the pairing gaps, listed in Table 1 and
drawn in Figs. \ref{B2-DELTA},\ref{GAM-DELTA},\ref{B3-DELTA},
mostly lie in this range of values, so that we may say that
the adopted strength for $V_0$ is reasonable. 

Another possible source of ambiguity in evaluating the pairing gaps is the
proton-neutron isoscalar pairings which are expected to play an important
role in the $N=Z$ nuclei (see, for example, \cite{SD97,MF00} and
references therein). We have assumed that such isoscalar pairings
are absent in the states under consideration. Although this assumption should 
be examined, there are some experimental indications\cite{MF00,Vo00}
that this may be a fairly good approximation. 
It is clear that we need a more systematic and detailed investigations,
both theoretical calculations and experimental explorations,
for a better understanding of the pairing correlations in the
proton-rich $N=Z$ nuclei in the $A=64-84$ region.

In order to examine the sensitivity of the calculated results 
to the strength $V_0$ of the pairing interaction, 
we have made a calculation of the potential energy curve 
about the oblate shape in \Se\ 
as a function of the triangular octupole deformation parameter 
$\beta_{33}$ for $V_0=-900,-1000$ and $-1100$ MeV$\cdot$fm$^3$. 
The result is shown in Fig. \ref{V-DEP}. 
As expected, the potential energy curve becomes shallower with
increasing (absolute value of) $V_0$. 
Thus, the local minimum at $\beta_{33}\approx 0.10$
disappears with 10\% increase of the (absolute) value of $V_0$.
In any case, the potential is so shallow that we cannot associate
a definite physical significance with the equilibrium values of $\beta_{33}$.  
We can still draw from these calculations an important conclusion that
the oblate ground state of \Se\ is extremely soft with respect to the
triangular octupole deformation.

\subsection{\it Discussions}

Actually, we need a more detailed investigation on the physical
implication of the extremely soft potentials like those with respect 
to the triangular deformation in \Se\ and for the tetrahedral
shape degree of freedom in \Zr. 
As is well known in the case of the axially symmetric $Y_{30}$
octupole deformation\cite{NA76,MF83,BK91,MB95,ER91}, a definite
minimum develops at finite value of $\beta_{30}$ after the parity
projection when the mean-field potential is very soft with respect to
$\beta_{30}$. 
For the case of non-axial octupole deformations,
a similar effect of the parity projection has been demonstrated by
Takami, Yabana and Ikeda\cite{TY96} for light nuclei.
It remains to be examined whether or not 
the situation is similar for the non-axial octupole deformations 
in medium-mass nuclei under consideration.

More generally speaking, investigations of modes of excitation
and of excitation spectra 
associated with the instabilities toward the non-axial
octupole shape deformations is one of the major challenges for future.
The present paper should be regarded as providing a HFB mean-field basis 
for a study of dynamics by means of methods 
like the quasiparticle RPA and the SCC method\cite{MM80}.

\section{Conclusion}

We have constructed a new computer code that carries out 
Skyrme-HFB calculations in the 3-dimensional Cartesian-mesh space 
without imposing any restriction on the spatial symmetry,
and investigated shape coexistence and non-axial octupole deformations 
in proton-rich $N=Z$ nuclei, \Ge, \Se, \Kr, \Sr, \Zr\ and \Mo.  
The ground state shape changes from triaxial (\Ge), 
oblate (\Se, \Kr), large prolate (\Sr, \Zr), 
to spherical (\Mo) as $N(=Z)$ increases,
in agreement with the available experimental data and 
the previous theoretical calculations. 
The extreme softness toward the \Ythree\ triangular deformation
of the oblate ground state of \Se\ and that toward 
\Ytwo\ tetrahedral deformation of the excited spherical minimum of \Zr,
pointed out by Takami et al.\cite{TY98,MT99} on the basis of the
Skyrme-HF plus BCS calculations, have been confirmed 
by the fully selfconsistent Skyrme-HFB calculations with the use of 
the density-dependent zero-range pairing interaction.

The symmetry-unrestricted Skyrme-HFB computer code constructed in this
work provides a selfconsistent mean-field basis for future investigation
of collective modes of excitation in neutron-rich nuclei with neutron skins
as well as in proton-rich nuclei.

\section*{Acknowledgements}

During the course of this work, we have benefited from discussions
with J. Dobaczewski, I. Hamamoto, P.-H. Heenen, S. Mizutori, W. Nazarewicz, 
H. Sagawa, N. Tajima, S. Takami, J. Terasaki, K. Yabana, and X.Z. Zhang.  
We would like to express our hearty thanks to them.
The numerical calculations were performed on the NEC SX-4 supercomputers
at Osaka University and at Yukawa Institute for Theoretical Physics, 
Kyoto University. 
This work was supported by the Grant-in-Aid for Scientific Research
(No. 10640264) from the Japan Society for the Promotion of Science.

\newpage

\begin{table*}%[b!]

\begin{tabular}{c||ccc}
\hline
         & Oblate  
         & Spherical 
         & Prolate  \\ 

         \hline \hline

         &   &   & g.s. \\

 $^{64}$Ge
         & 
         &
         &  $\beta ,\gamma = 0.27, 25^{\circ}$ (triaxial) \\

         & 
         &
         &  $\beta_{3} = 0.0$  \\ 
 
         & 
         &
         &  $\Delta_p = 1.25, \Delta_n = 1.12$  \\ \hline

         &  g.s. &   & 0.52 \\

 $^{68}$Se
         & $\beta ,\gamma  = 0.28, 60^{\circ}$ 
         & 
         & $\beta ,\gamma = 0.26, 0^{\circ}$ \\

         &  $\beta_{3} = \beta_{33} \approx 0.08$ 
         &  
         &  $\beta_{3} = 0.0$   \\ 

         & $\Delta_p = 1.28, \Delta_n = 1.13$
         &
         & $\Delta_p = 1.29, \Delta_n = 1.15$  \\ \hline

         & g.s.   & & 0.92 \\
 $^{72}$Kr
         & $\beta ,\gamma = 0.32, 60^{\circ}$
         & 
         & $\beta ,\gamma = 0.40, 0^{\circ}$ \\

         &  $\beta_{3} = 0.0$
         &  
         &  $\beta_{3} = 0.0$  \\ 

         & $\Delta_p = 1.03, \Delta_n = 1.23$
         &
         & $\Delta_p = 1.25, \Delta_n = 0.92$  \\ \hline

         &  1.79 &  & g.s. \\
 $^{76}$Sr
         & $\beta ,\gamma  = 0.30, 60^{\circ}$ 
         &  
         & $\beta ,\gamma = 0.51, 0^{\circ}$ \\

         &  $\beta_{3} = \beta_{33} \approx 0.0$
         & 
         &  $\beta_{3} = 0.0$ \\ 

         & $\Delta_p = 1.47, \Delta_n = 1.43$
         &
         & $\Delta_p = 0.67, \Delta_n = 0.50$  \\ \hline

         &  0.86 & 1.01 & g.s. \\
 $^{80}$Zr
         & $\beta ,\gamma = 0.20, 60^{\circ}$ 
         & $\beta ,\gamma = 0.0, 0^{\circ}$
         & $\beta ,\gamma = 0.51, 0^{\circ}$ \\
 
         &  $\beta_{3} = 0.0$ 
         &  $\beta_{3} = \beta_{32} \approx 0.15 $
         &  $\beta_{3} = 0.0$ \\ 

         & $\Delta_p = 1.02, \Delta_n = 0.82$
         & $\Delta_p = 0.68, \Delta_n = 0.39$
         & $\Delta_p = 0.79, \Delta_n = 0.78$  \\ \hline

         &  0.20 & g.s. & 1.52 \\
 $^{84}$Mo
         & $\beta ,\gamma = 0.16, 60^{\circ}$ 
         & $\beta ,\gamma = 0.0, 0^{\circ}$
         & $\beta ,\gamma = 0.66, 0^{\circ}$ \\
 
         &  $\beta_{3} = 0.0$ 
         &  $\beta_{3} = \beta_{30} \approx 0.0 $
         &  $\beta_{3} = 0.0$ \\ 

         & $\Delta_p = 1.46, \Delta_n = 1.42$
         & $\Delta_p = 0.74, \Delta_n = 0.72$
         & $\Delta_p = 0.0, \Delta_n = 0.0$  \\ \hline

\end{tabular}
\caption{
\small Solutions of the HFB equations for proton-rich $N=Z$ nuclei
in the $A=64-84$ region.  For each nucleus, numbers in the first line 
indicate excitation energies measured from the ground state. 
The symbol $\approx$ indicates that the potential energy curve 
is extremely shallow about the equilibrium value.
Pairing gaps $\Delta_p$ and $\Delta_n$ are here defined as averages of
diagonal elements $\Delta_{i\bar{i}}$ over 5 MeV interval around
the Fermi surface, and their values (in MeV) at the equilibrium
deformations are listed.
}
\label{SUM_LM}
\end{table*}

\newpage

\newpage
\begin{figure}
\epsfxsize=13cm
\centerline{\epsffile{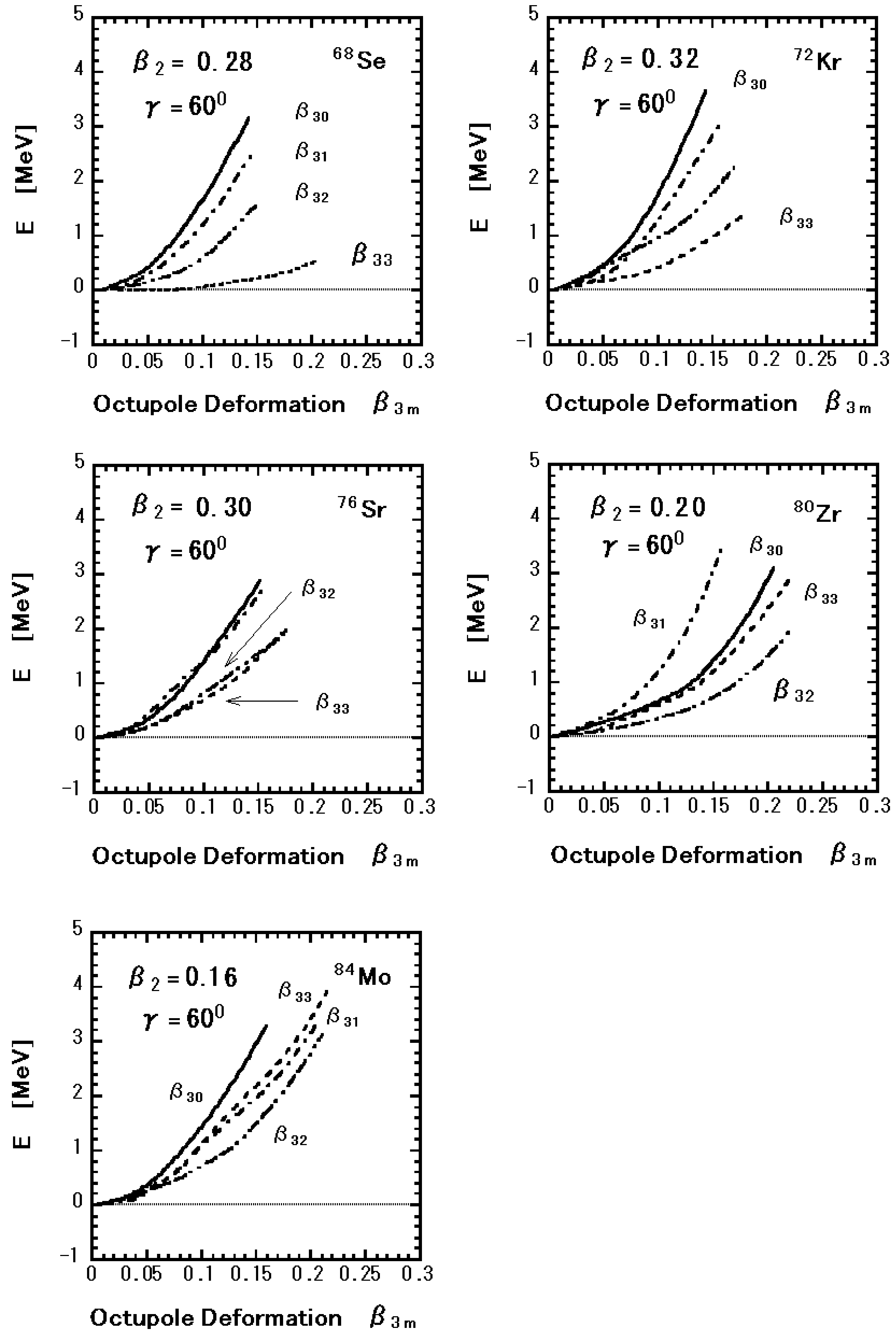}}
\caption{
\small Potential energy curves calculated by the constrained 
Skyrme-HFB method are drawn as functions of 
the octupole deformation parameters  $\beta_{3m}(m=0,1,2,3)$  
about the oblate minima (seen in Fig. 1) of the quadrupole deformation 
energies. 
One of the $\beta_{3m}(m=0,1,2,3)$ is varied while the other
$\beta_{3m}$'s are fixed to zero.
The effective interactions used are the same as in Fig. \ref{B2-E}.
}
%\vspace*{10pt}
\label{B3-E_ob}
\end{figure}

\begin{figure}
\epsfxsize=13cm
\centerline{\epsffile{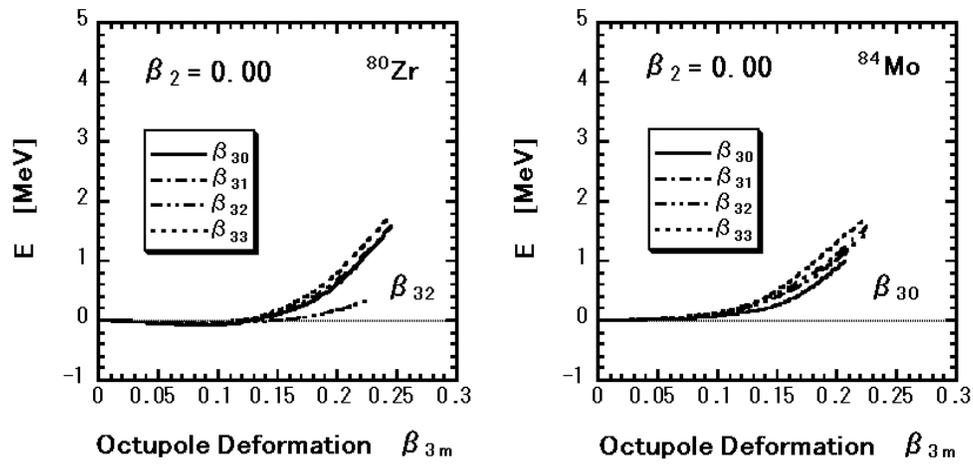}}
\caption{
\small Same as Fig. \ref{B3-E_ob} but about the spherical minima.
}
%\vspace*{10pt}
\label{B3-E_sph}
\end{figure}

\begin{figure}
\epsfxsize=13cm
\centerline{\epsffile{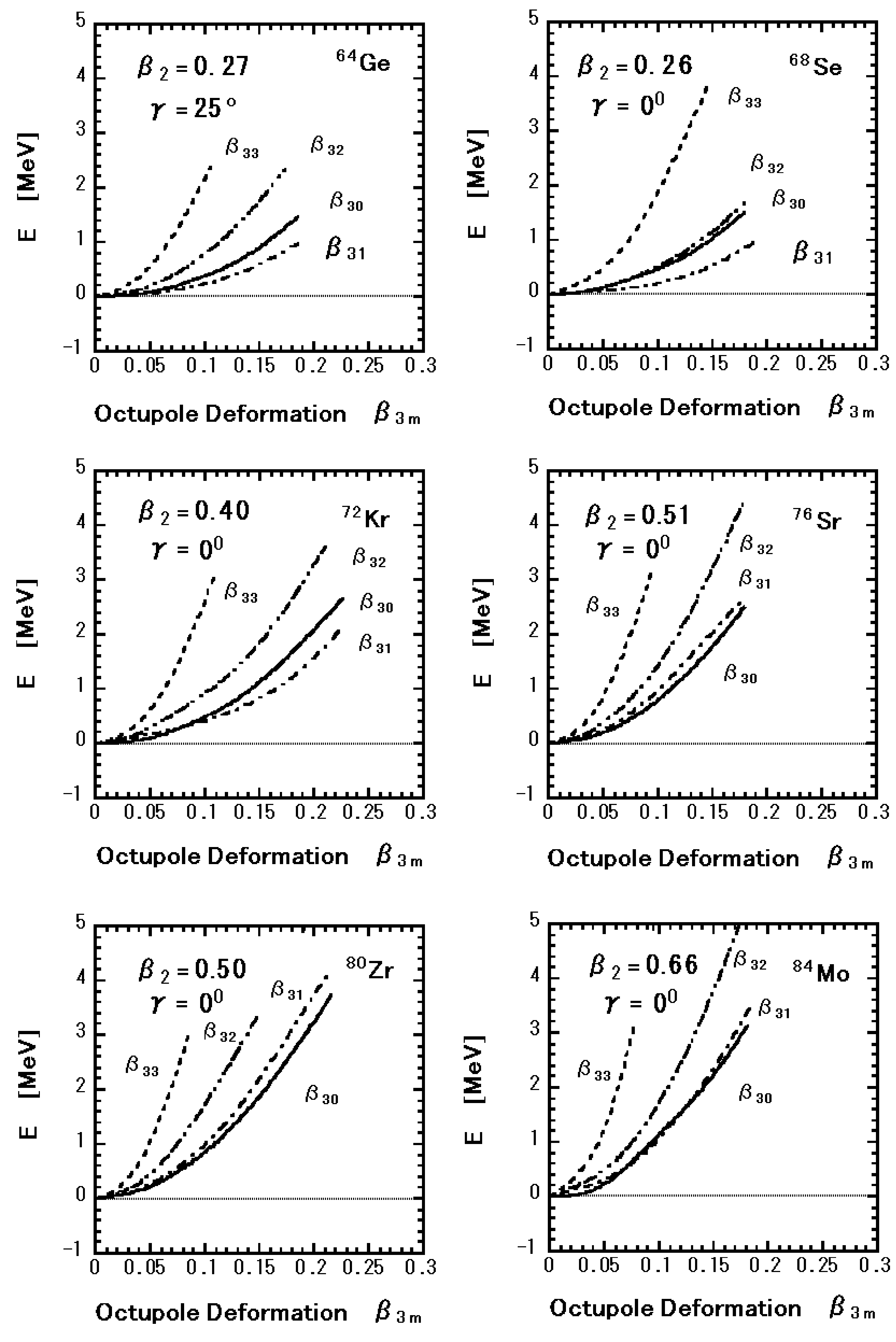}}
\caption{
\small Same as Fig. \ref{B3-E_ob} but about the prolate minima
(the triaxial minimum in the case of \Ge).
}
%\vspace*{10pt}
\label{B3-E_pr}
\end{figure}

\begin{figure}
\epsfxsize=13cm
\centerline{\epsffile{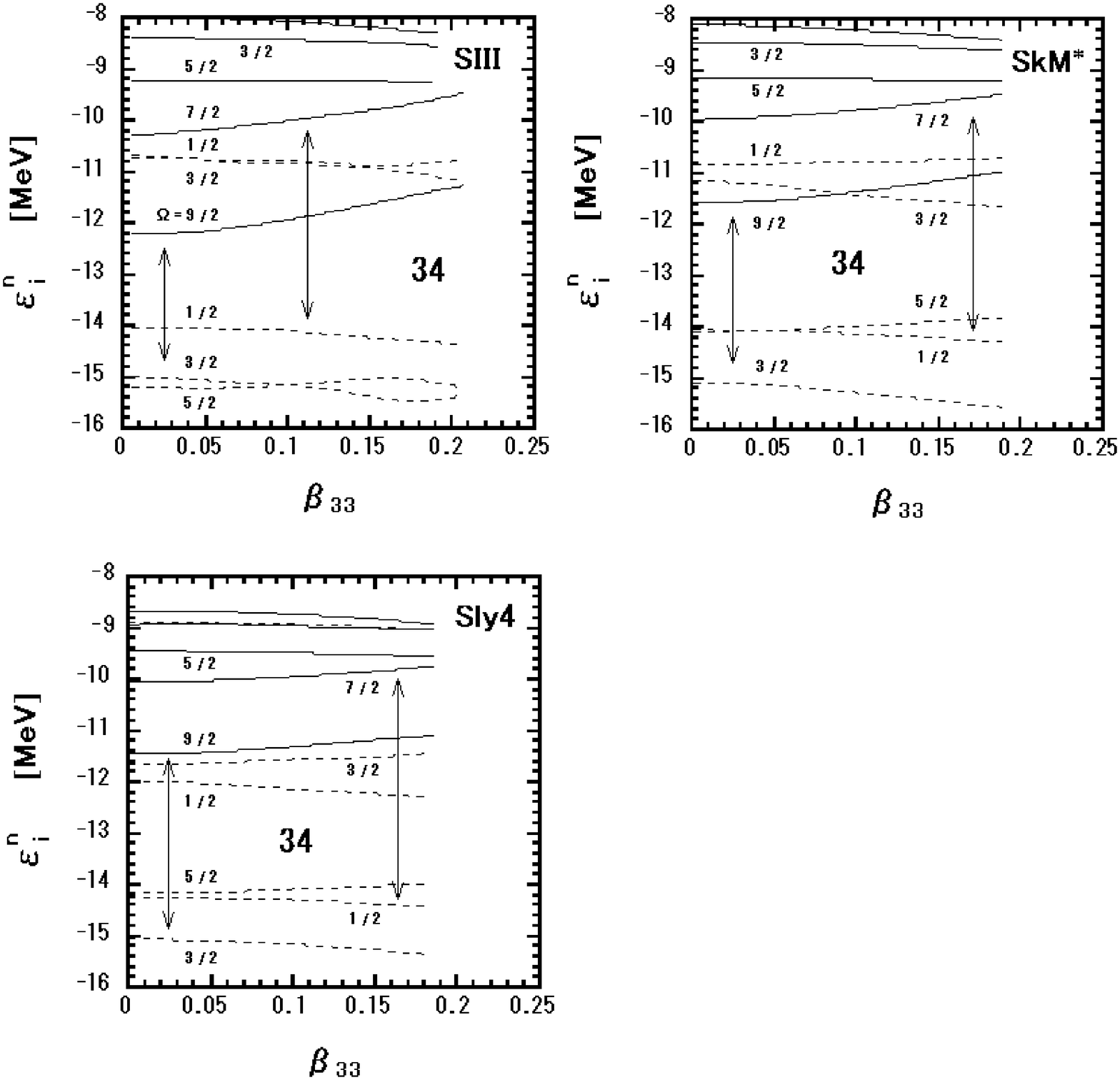}}
\caption{
\small Neutron single-particle energies for \Se\ plotted as functions of 
the octupole deformation parameter $\beta_{33}$ about the oblate shape. 
Here, the single-particle energies mean eigenvalues of the HF
Hamiltonian with  the density $\rho\left( \boldr \right)$
determined by the HFB equations.
Results for the SIII, SkM$^*$ and SLy4 parameter sets are compared.
Equilibrium quadrupole deformations obtained for each Skyrme interaction are 
$\beta_2 = -0.28, -0.25$ and $-0.24$ for SIII, SkM$^*$ and SLy4, respectively.
Solid (broken) lines indicate levels which have positive (negative) parity
in the limit $\beta_{33}=0$.
The projection of the angular momentum on the symmetry axis, $\Omega$, is
a good quantum number only at $\beta_{33}=0$.
The arrows indicate the $\Delta\Omega=3$ coupling associated with the 
triangular \Ythree\ deformation as discussed in the text.
The single-particle spectrum for protons is almost the same as for neutrons.
}
%\vspace*{10pt}
\label{SPE-SE}
\end{figure}

\begin{figure}
\epsfxsize=13cm
\centerline{\epsffile{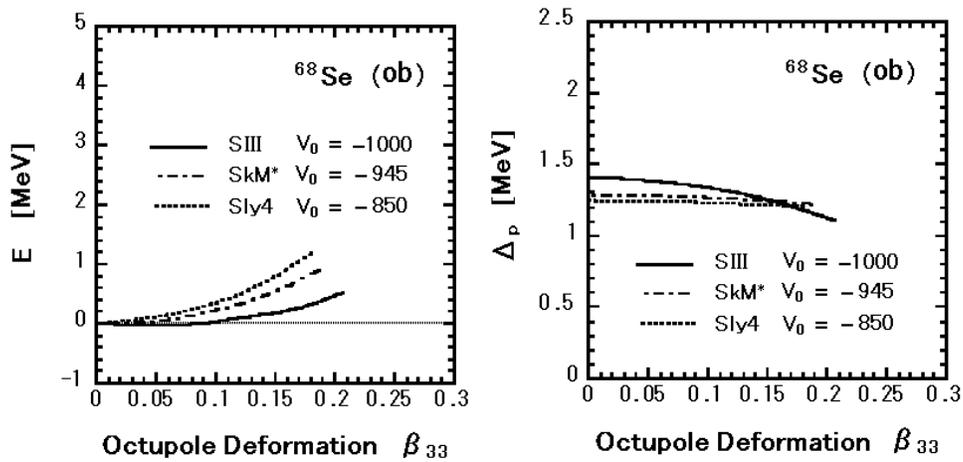}}
\caption{
\small Comparison of the HFB potential energy curves for \Se\ 
about the oblate shape  
as functions of the triangular deformation parameter $\beta_{33}$, 
calculated for different versions of the Skyrme interaction
(left-hand side). 
The pairing-interaction strengths $V_0$ are chosen
such that the average pairing gaps become approximately equal for
all Skyrme interactions (as displayed in the right-hand side). 
The calculated deformation parameter $\beta_2$ are 
$-0.28, -0.25$ and $-0.24$ for the SIII, SkM$^*$ and SLy4 interactions, 
respectively.
}
%\vspace*{10pt}
\label{SK-DEP}
\end{figure}

\begin{figure}
\epsfxsize=13cm
\centerline{\epsffile{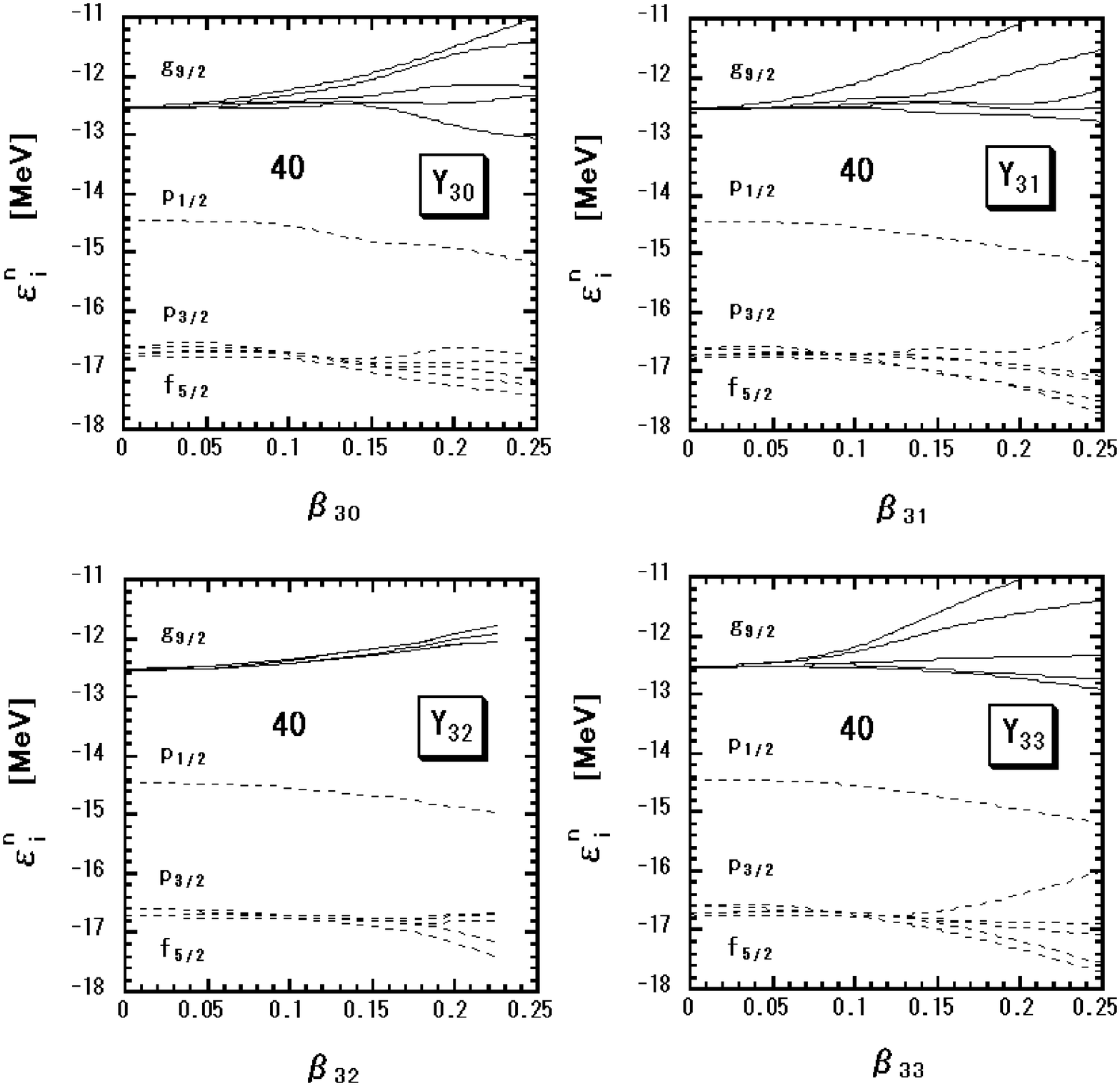}}
\caption{
\small Neutron single-particle energies for \Zr\ plotted as functions of 
the octupole deformation parameters $\beta_{3m}  (m = 0, 1, 2, 3)$ 
about the spherical shape. 
Here, the single-particle energies mean eigenvalues of the HF
Hamiltonian with  the density $\rho\left( \boldr \right)$
determined by the HFB equations.
The SIII interaction is used.
Solid (broken) lines indicate levels which have positive (negative) parity
in the limit $\beta_{32}=0$.
The single-particle spectrum for protons is almost the same as for neutrons.
}
%\vspace*{10pt}
\label{SPE-ZR}
\end{figure}

\begin{figure}
\epsfxsize=8cm
\centerline{\epsffile{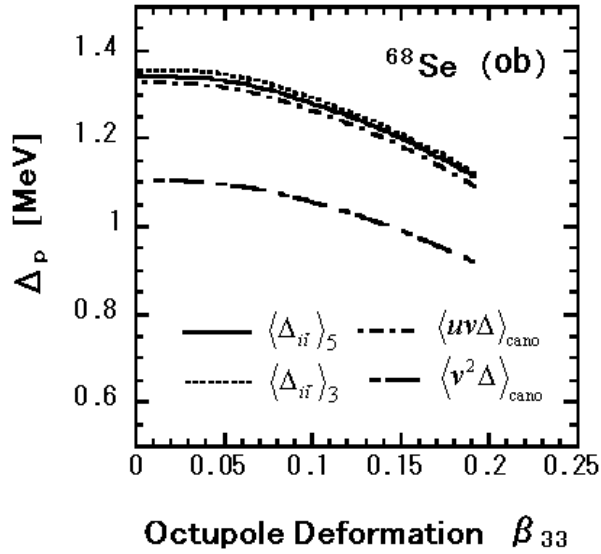}}
\caption{
\small Comparison of differently defined average pairing gaps 
for \Se, plotted as functions of the triangular deformation 
parameter $\beta_{33}$ superposed on the oblate shape. 
Here,
$
\left\langle {\Delta _{i\bar i}} \right\rangle _{\Delta E}
={{\sum_i {f_if_{\bar i}g_ig_{\bar i}\Delta _{i\bar i}}} 
/{\sum_i {f_if_{\bar i}g_ig_{\bar i}}}} 
$,
$
\left\langle uv\Delta  \right\rangle _{cano}=
\sum_\alpha  u_\alpha v_\alpha \left\langle
\varphi _\alpha \left| \Delta  \right|\varphi _\alpha 
\right\rangle  / \sum_\alpha  u_\alpha v_\alpha 
$\cite{TO94,BR00,DB00}
and
$
\left\langle v^2\Delta  \right\rangle _{cano}=
\sum_\alpha  v_\alpha ^2\left\langle
\varphi _\alpha \left| \Delta  \right|\varphi _\alpha 
\right\rangle  / \sum_\alpha  v_\alpha ^2
$\cite{DN96b},
where $f_i=( 1+\exp[ (\varepsilon _i-\lambda_F-\Delta E/2)/ \mu])^{-1/4},
~g_i=( 1+\exp[ (\varepsilon _i-\lambda_F+\Delta E/2)/ \mu ])^{-1/4} 
$
with $\Delta E=$ 3 or 5 MeV. 
}
%\vspace*{10pt}
\label{DELTA-DEF}
\end{figure}

\begin{figure}
\epsfxsize=13cm
\centerline{\epsffile{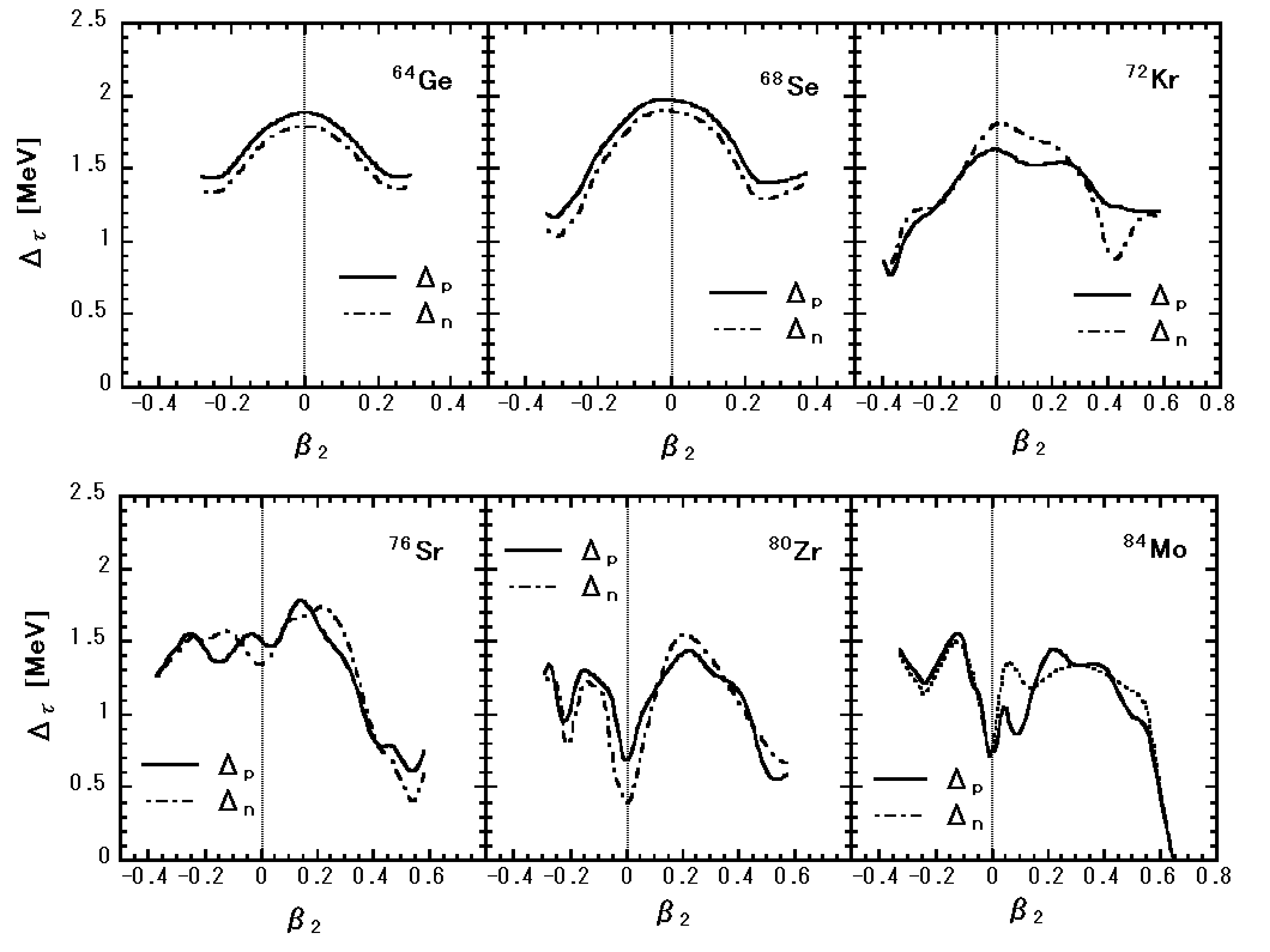}}
\caption{
\small Variations of the pairing gaps $\Delta_{\tau}$ ($\tau$= p, n) 
calculated by the constrained Skyrme-HFB method
as functions of the quadrupole deformation 
parameter $\beta_2$ for \Ge, \Se, \Kr, \Sr, \Zr\ and \Mo. 
The effective interactions used are the same as in Fig. \ref{B2-E}.
}
%\vspace*{10pt}
\label{B2-DELTA}
\end{figure}

\begin{figure}
\epsfxsize=13cm
\centerline{\epsffile{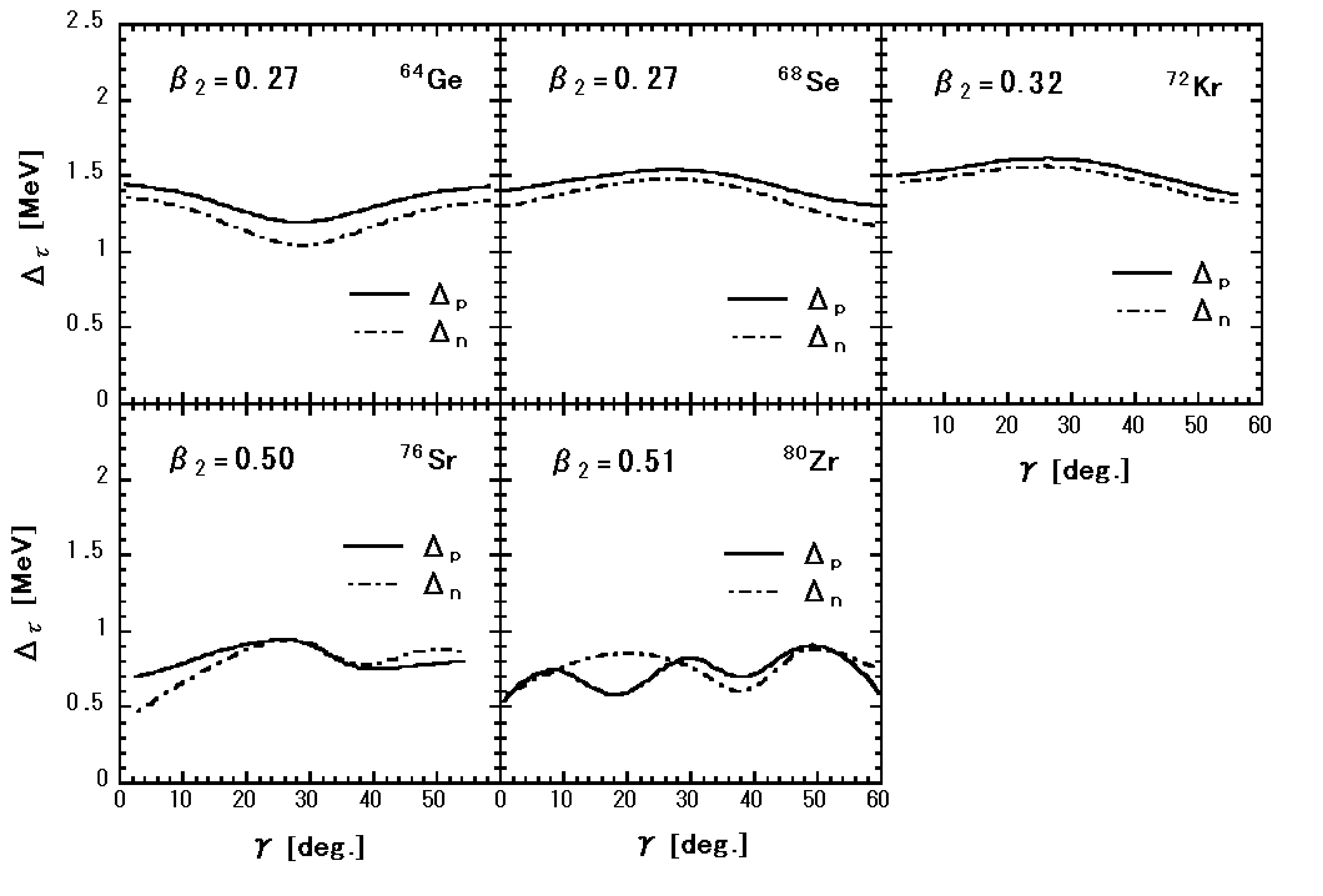}}
\caption{
\small Variations of the pairing gaps $\Delta_{\tau}$ ($\tau$ = p, n) 
calculated by the constrained Skyrme-HFB method
as functions of the triaxial deformation  
parameter $\gamma$ at fixed $\beta_2$ for \Ge, \Se, \Kr, \Sr\ and \Zr.
The effective interactions used are the same as in Fig. \ref{B2-E}.
}
%\vspace*{10pt}
\label{GAM-DELTA}
\end{figure}

\begin{figure}
\epsfxsize=13cm
\centerline{\epsffile{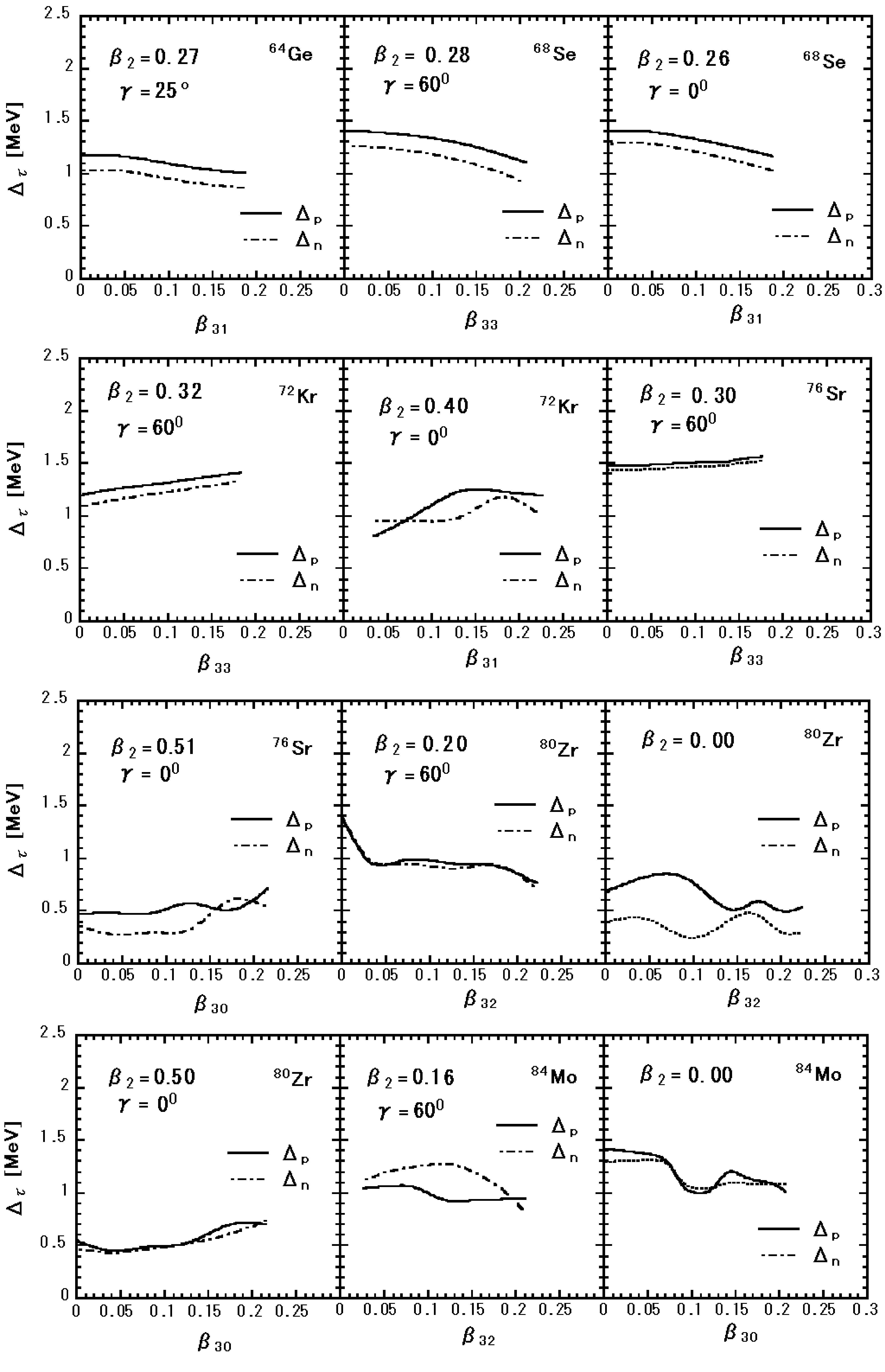}}
\caption{
\small Variations of the pairing gaps $\Delta_{\tau}$ ($\tau$= p, n) 
calculated by the constrained Skyrme-HFB method
as functions of the octupole deformation  
parameter $\beta_{3m}  (m = 0, 1, 2, 3)$ about the local minima
(seen in Fig. 1) of the quadrupole deformation energies
for \Ge, \Se, \Kr, \Sr, \Zr\ and \Mo.
The effective interactions used are the same as in Fig. \ref{B2-E}.
}
%\vspace*{10pt}
\label{B3-DELTA}
\end{figure}

\begin{figure}
\epsfxsize=13cm
\centerline{\epsffile{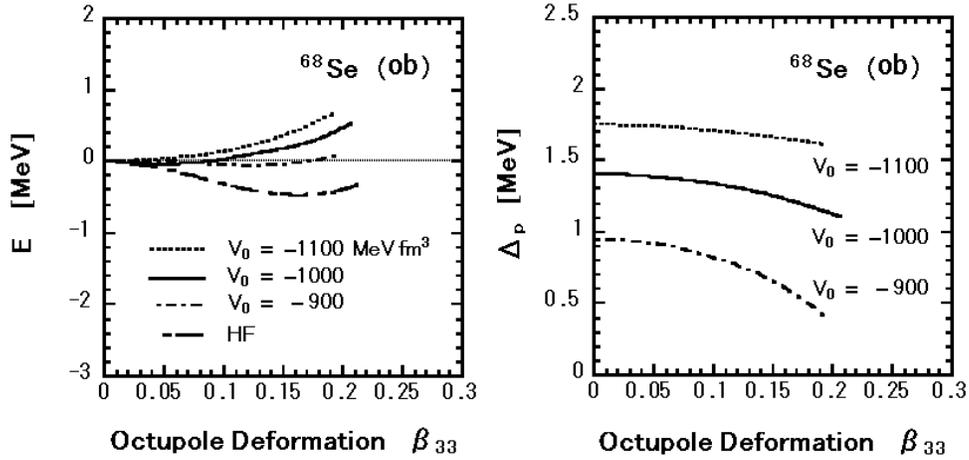}}
\caption{
\small Comparison of the potential energy curves (left-hand side)
and average pairing gaps for protons (right-hand side) 
calculated by the constrained Skyrme-HFB method 
as functions of the triangular deformation parameter $\beta_{33}$ 
about the oblate shape for \Se\ with use of 
different strengths $V_0$ of the density-dependent pairing interaction
(and with the same SIII interaction ). 
}
%\vspace*{10pt}
\label{V-DEP}
\end{figure}

\end{document}